\documentclass[preprint,showpacs,floatfix,pre]{revtex4}
\usepackage{amsmath,amssymb}
\usepackage[dvips]{graphicx}

\begin{document}
\title{The Rubber Band Revisited: Wang-Landau Simulation}
\author{Lucas S. Ferreira}
\email{hotwister@hotmail.com}

\author{Alvaro A. Caparica}
\email{caparica@if.ufg.br}
\affiliation{Departamento de F\'{\i}sica, Universidade Federal do Goi\'{a}s, C.P. 131,
74001-970 Goi\^{a}nia-GO, Brazil}

\author{Minos A. Neto}
\email{minos@pq.cnpq.br}

\author{Mircea D. Galiceanu}
\email{mircea@ufam.edu.br}

\affiliation{Departamento de F\'{\i}sica, Universidade Federal do Amazonas, 3000, Japiim,
69077-000, Manaus-AM, Brazil}
\date{\today}

\begin{abstract}

In this work we apply Wang-Landau simulations to a simple model which has exact solutions both in the microcanonical and canonical formalisms.
The simulations were carried out by using an updated version of the Wang-Landau sampling. We consider a homopolymer chain consisting of $N$
monomers units which may assume any configuration on the two-dimensional lattice. By imposing constraints to the moves of the polymers we
obtain three different models. Our results show that updating the density of states only after every $N$ monomers moves leads to a better
precision. We obtain the specific heat and the end-to-end distance per monomer and test the precision of our simulations comparing the location
of the maximum of the specific heat with the exact results for the three types of walks.

\textbf{Keywords}: Homopolymer, Monte Carlo, Wang-Landau

\end{abstract}

\maketitle

\section{Introduction\protect\nolinebreak}

The sequencing of the Human Reference Genome, announced ten years ago, provided a roadmap that is the basic foundation for modern biomedical research
\cite{elaine}. This monumental achievement was enabled by developments in DNA (homopolymer) sequencing technology that allowed data production which
exceed the original descripion of Sanger sequencing \cite{sanger}. Linear polymers are the simplest physical systems that can be studied in the
framework of random walks models. They are long chain-like molecules formed by repetition of a basic unit or segment, where more importantly the polymer
is \textit{flexible}, i.e., it can assume different geometric configurations.

Recently, the study of homopolymers has been established by various techniques in condensed matter physics. Cohen \textit{et al.} \cite{cohen}, studied
the behavior of single file translocation of a homopolymer through an active channel under the presence of a driving force by using Langevin dynamics
simulation. Previous works on homopolymers which studied the denaturation of circular DNA are extensions of the Poland-Scheraga model \cite{bar}.
Experimentally, the viscoelastic properties of a binary mixture of a mesogenic side-chain block copolymer in a low molecular weight nematic liquid
crystal are studied for mass concentrations ranging from the diluted regime up to a liquid crystalline gel state \cite{maxim}.

Although Monte Carlo simulations play an important role for the study of phase transitions and critical phenomena, some well-known difficulties arise when
one uses standard algorithms (one-flip algorithms) \cite{metropolis} for the study of random walks models. These difficulties have been overcomed by the
development of alternative Monte Carlo methods, such as parallel-tempering \cite{nemoto}, cluster algorithms \cite{wolff}, multicanonical algorithms
\cite{berg}, and more recently the Wang-Landau method \cite{wanglandau}. This method has been applied with great success to many systems, in particular to
polymers in lattice \cite{cunha, vorontsov, binder}.

In the present paper, using the Wang-Landau method, we investigate the computer simulations of a homopolymer model with exact solution in canonical and
microcanonical formalism. In section \ref{models} we give an introduction to the three studied models and we briefly present the mathematical background. In section \ref{simulations} we outline shortly how the simulations were carried out.
In section \ref{results} we show and discuss the results for all the models.

\section{Models and Formalism}\label{models}

\subsection{Model 1}\label{model1}

\begin{figure}[htbp]
\centering
\includegraphics[width=7.6cm,height=8.9cm]{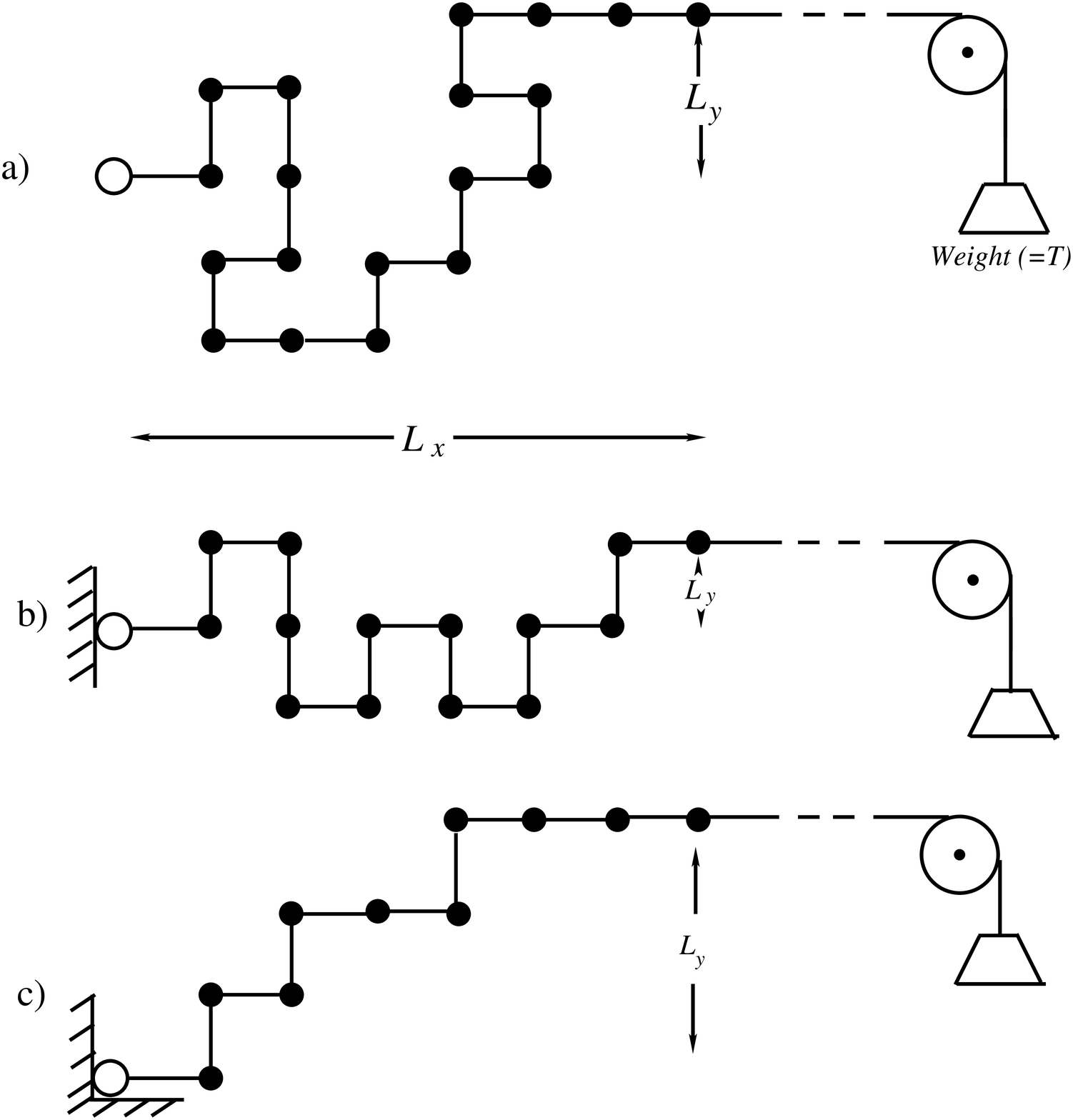}
\caption{Here we consider three types of walks: (a) Model 1: unrestricted motion along the $x$-axis and $y$-axis, (b) Model 2: is forbidden the backward
motion on the $x$-axis, and (c) Model 3: the random walk moves only along the positive directions of $x$-axis and $y$-axis.}
\label{model}
\end{figure}

We consider a homopolymer chain of $N$ monomers units of length $a$ which may assume any configuration on a two-dimensional latice \cite{callen}. One end of
the polymer is fixed and it´s taken as the origin of coordinates, shown by an open circle in Figure \ref{model}, a). The other end of this linear chain is subject to an externally applied tension $\tau$, acting
along the positive $x$-axis. A possible realization of this model is sketched in Figure \ref{model}, a). Each polymer unit is permitted to lie either parallel or antiparallel
to the $x$-axis and we assign the works $-\tau a$ and $+\tau a$ to these two orientations. We denote by $L_x$ the distance between the ends of the polymer chain,
on the $x$-direction. Each monomer unit has the additional possibility of lying perpendicular to the $x$-axis, in the $+y$ or $-y$ directions. We associate a
positive energy $\varepsilon$ to such a perpendicular monomer and the distance between the ends of the chain on the $y$-direction is denoted by $L_y$.

The hamiltonian of this model can be written as
\begin{equation}\label{h_1}
\mathcal{H}_1=\left(N^{+}_{y}+N^{-}_{y}\right)\varepsilon+\left(N^{-}_{x}-N^{+}_{x}\right)\tau a,
\end{equation}
where $N^{+}_{x}$ and $N^{-}_{x}$ are the number of monomers along the $+x$ and $-x$ directions respectively, and similarly for $N^{+}_{y}$ and $N^{-}_{y}$. Since
the tension in the $y$ direction is zero, we can assume $N^{+}_{y}=N^{-}_{y}$. Then
\begin{equation}\label{N_1}
N=N^{+}_{x}+N^{-}_{x}+N^{+}_{y}+N^{-}_{y},
\end{equation}

\begin{equation}\label{Lx_1}
N^{+}_{x}-N^{-}_{x}=L_{x},
\end{equation}

\begin{equation}\label{U_1}
N^{+}_{y}+N^{-}_{y}=U,
\end{equation}
from which we find

\begin{equation}\label{Nx_1}
 N^{+}_{x}=\frac{1}{2}\left(N-U+L_{x}\right),
\end{equation}

\begin{equation}\label{Nx1_1}
N^{-}_{x}=\frac{1}{2}\left(N-U-L_{x}\right),
\end{equation}
and

\begin{equation}\label{Ny_1}
N^{+}_{y}=N^{-}_{y}=\frac{1}{2} U.
\end{equation}

The number of configurations of the polymer consistent with a given end-to-end distance in the $x$-direction, $L_{x}$ (the dimensionless length of the polymer), and $U$ (the number of monomers lying perpendicular
to the $x$-axis), is

\begin{equation}\label{omega_1}
\Omega\left(L_{x},U,N\right)=\frac{N!}{N^{+}_{x}!N^{-}_{x}!N^{+}_{y}!N^{-}_{y}!}.
\end{equation}

From Eq. (\ref{h_1}), if we set $\varepsilon \equiv \tau a = 1$, we can write for a given energy level

\begin{equation}\label{E_1}
E= U-N^{+}_{x}+N^{-}_{x},
\end{equation}
and from Eq. (\ref{N_1})
\begin{equation}\label{N1_1}
N=U+N^{+}_{x}+N^{-}_{x}.
\end{equation}

Adding the equations (\ref{E_1}) and (\ref{N1_1}) we obtain
\begin{equation}\label{U_Nx_1}
U+N^{-}_{x}=\frac{N+E}{2}.
\end{equation}

Inserting Eqs. (\ref{Nx_1}), (\ref{Nx1_1}) and (\ref{Ny_1}) into (\ref{omega_1}), using Eqs. (\ref{E_1}), (\ref{N1_1}), and (\ref{U_Nx_1}), and setting
$N^{-}_{x}\equiv n$ we obtain the number of configurations with energy $E$ as

\begin{equation}\label{g_1}
g(E)=\sum_{u=0}^{\frac{N+E}{2}} \sum_{n=0}^{u} \frac{N!}{\left(\frac{N-E}{2}\right)!\left(\frac{N+E}{2}-u\right)!n!(u-n)!}.
\end{equation}

Using the definition of the entropy and defining $\partial S/\partial U=1/T$, we obtain \cite{callen}:

\begin{equation}\label{LxN_1}
\frac{L_{x}}{N}=\frac{N \sinh{(\tau a/k_BT)}}{\cosh{(\tau a/k_BT)}1+\exp\left( -\varepsilon /k_{B}T \right)}.
\end{equation}

\subsection{Model 2}\label{model2}

In this model we consider that each polymer unit is parallel to the $x$-axis; no antiparallel move is allowed. Additionally the polymer units have the
possibility of lying on the $+y$ and $-y$ directions. In Figure \ref{model} b) is sketched a possible configuration of this model, where we depicted by an impenetrable wall the forbidden region along the $x$-axis.

The hamiltonian for the model $2$ can be written as
\begin{equation}\label{H_2}
\mathcal{H}_2=\left(N^{+}_{y}+N^{-}_{y}\right)\varepsilon-N^{+}_{x}\tau a,
\end{equation}
where $N_{x}^{+}$ is the number of monomers along the $+x$ direction and similarly for $N_{y}^{+}$ and $N_{y}^{-}$. Assuming
again $N_{y}^{+}\equiv N_{y}^{-}$ we obtain the following equations:

\begin{equation}\label{Nx_2}
 N^{+}_{x}=\frac{1}{2}\left(N-U+L_{x}\right),
\end{equation}

and

\begin{equation}\label{N+y_2}
N^{+}_{y}=N^{+}_{y}=\frac{U}{2},
\end{equation}

For this model the number of configurations of the polymer consistent with a given $L_{x}$ and $U$ is

\begin{equation}\label{omega_2}
\Omega\left(L_{x},U,N\right)=\frac{N!}{N^{+}_{x}!N^{+}_{y}!N^{-}_{y}!}.
\end{equation}
From Eq. (\ref{H_2}), if we set $\varepsilon \equiv \tau a = 1$, we can write the energy as

\begin{equation}\label{E_2}
E= U-N^{+}_{x},
\end{equation}
and for $N=N^{+}_{y}+N^{-}_{y}+N^{+}_{x}$ we obtain
\begin{equation}\label{N_2}
N=U+N^{+}_{x}.
\end{equation}

After similar calculations as for model $1$, subsection \ref{model1}, we obtain the density state and end-to-end distance per monomer, respectively

\begin{equation}\label{g_2}
g(E)=\sum_{n=0}^{\frac{N+E}{2}}\frac{N!}{\left(\frac{N-E}{2}\right)!\left(\frac{N+E}{2}-n\right)!n!}
\end{equation}%
and
\begin{equation}\label{LxN_2}
\frac{L_{x}}{N}=\frac{1}{1+2\exp\left( - \frac{2\tau a}{k_{B}T}\right)}.
\end{equation}

\subsection{Model 3}\label{model3}

In this model we consider that each polymer unit is parallel to the $x$-axis and we allow only the possibility of lying on the $+y$ direction. Thus, we restrict
to the situation of positive values for both axes. In Figure \ref{model} c) we show a possible configuration of this model.

The hamiltonian for the model can be written as
\begin{equation}\label{H_3}
\mathcal{H}_3=N^{+}_{y}\varepsilon-N^{+}_{x}\tau a,
\end{equation}
where $N_{x}^{+}$ is the number of monomers along the $+x$ direction and similarly for $N_{y}^{+}$. Then

Using the method previously described, we obtain $N_{x}^{+}$ and $N_{y}^{+}$ for this model

\begin{equation}\label{Nx_3}
N^{+}_{x}=\frac{1}{2}\left(N-U+L_{x}\right)
\end{equation}
and
\begin{equation}\label{Ny_3}
N^{+}_{y}=\frac{1}{2}\left(N+U-L_{x}\right).
\end{equation}

For this model the number of configurations with energy $E$ is given by

\begin{equation}\label{g_3}
g(E)=\frac{N}{\left(\frac{N-E}{2}\right)!\left(\frac{N+E}{2}\right)!}.
\end{equation}

Using the equation for $g(E)$ we obtain the length of the polymer as

\begin{equation}\label{LxN_3}
\frac{L_{x}}{N}=\frac{1}{1+\exp\left( - \frac{2\tau a}{k_{B}T}\right)}.
\end{equation}

\section{Simulations}\label{simulations}

In our simulations we followed the prescriptions of Ref. \cite{caparica1}. We define a Monte Carlo step (MCS) as giving sequentially to any unit the possibility
of changing its direction with identical probability to any allowed direction or remaining in the same one. At the beginning of the simulation we set $S(E)=0$
for all energy levels, where $S(E)\equiv\ln g(E)$. The random walk in the energy space runs through all energy levels from $E_{min}$ to $E_{max}$ with a probability
\begin{equation}\label{prob}
p(E\rightarrow E^{'})=\min\left\lbrace \exp\left[ \left( S(E)-S(E^{'})\right)\right],1\right\rbrace ,
\end{equation}
where $E$ and $E'$ are the energies of the current and the new possible configurations. After $N$ trial moves we update $H(E)\rightarrow H(E)+1$ and
$S(E)\rightarrow S(E)+F_{i}$, where $F_{i}=\ln f_{i}$, $f_{0}\equiv e=2.71828...$ and $f_{i+1}=\sqrt{f_{i}}$ (where $f_{i}$ is the so-called modification factor
and $H(E)$ is a histogram accumulated for each $f_i$). The flatness of the histogram is checked after a number of Monte Carlo (MC) steps and usually the histogram
is considered flat if $H(E)>0.8\langle H \rangle$, for all energies, where $\langle H \rangle$ is an average over the energies. If the flatness condition is
fulfilled we update the modification factor to a finer one and reset the histogram $H(E)=0$. The simulations are continued up to $f_{final}=f_{14}$ and the microcanonical
averages were accumulated from the very beginning ($f_{micro}=f_0$), results obtained by Ferrera and Caparica \cite{caparica2}. Having in hand the density of states,
one can calculate the canonical average of any thermodynamic variable as

\begin{equation}\label{mean}
\langle X\rangle_T=\dfrac{\sum_E \langle X\rangle_E g(E) e^{-\beta E}}{\sum_E g(E) e^{-\beta E}} ,
\end{equation}
where $\langle X\rangle_E$ is the microcanonical average accumulated during the simulations and $\beta=1/k_BT$,
$k_B$ is the Boltzmann constant and $T$ is the temperature.

\section{Results and Discussion}\label{results}

In Figure \ref{model} we depicted the three models which were presented in section \ref{models}. The difference between the models is given by the allowed moves. In the
first model, denoted by a) in the figure, we allow the polymer unit to move along the positive or the negative directions of the $x$-axis, the same for the $y$-axis. The
end-to-end distance of the polymer on $x$-direction is denoted by $L_x$ and with $L_y$ we denote this distance on $y$-direction. In the second model, b) in figure
\ref{model},  the antiparallel motion along the $x$-axis is forbidden. In the third considered model there are allowed  moves only on the positive side of both $x$ and $y$-axis.

Using the simulated and the exact density of states in Eq.\eqref{mean} we calculate the specific heat given by
\begin{equation}\label{heat}
 C=\frac{\langle(E-\langle E\rangle)^2\rangle}{T^{2}}
\end{equation}
and the mean end-to-end distance
\begin{equation}\label{distance}
 \langle L_x\rangle=\langle|x_N-x_1|\rangle,
\end{equation}
where $E$ is the energy of the configurations and $x_1$ and $x_N$ are the corresponding $x$-coordinates of the
ends of the polymer.

\begin{figure}[htbp]
\centering
\includegraphics[width=7.5cm,height=7.5cm]{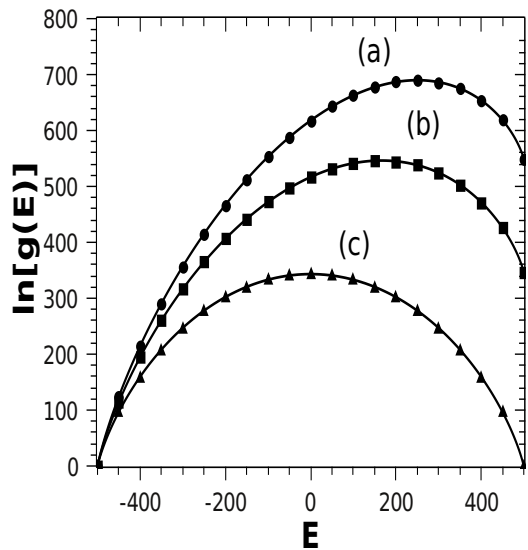}
\caption{Exact density of states $g(E)$ for polymer size $N=500$ for (a) Model 1; (b) Model 2 and (c) Model 3. The lines are the exacts results and the
symbols are simulational results.}
\label{ge}
\end{figure}

\begin{figure}[htbp]
\centering
\includegraphics[width=7.5cm,height=7.5cm]{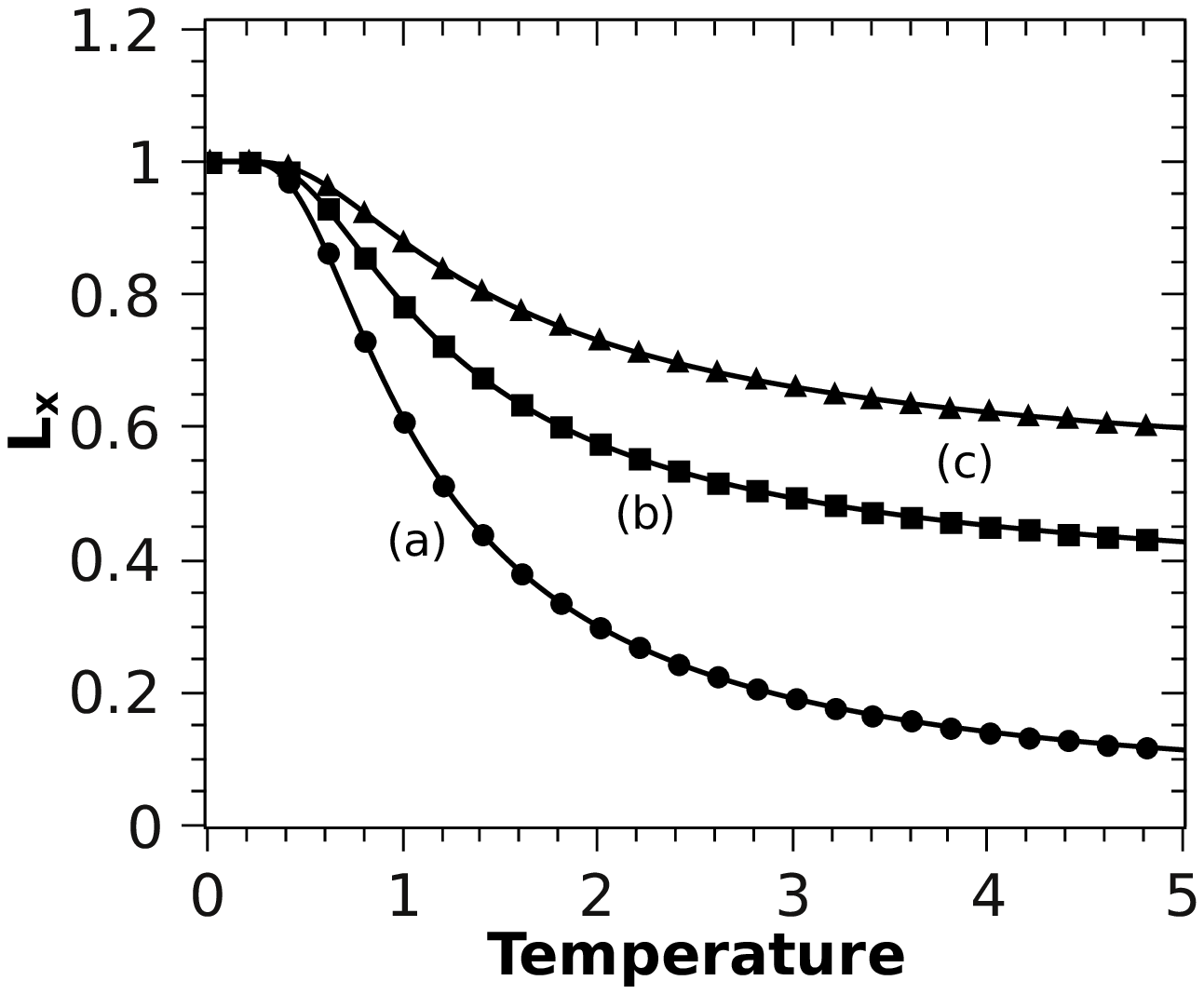}
\caption{End-to-end distance per monomer for polymer size $N=500$ for (a) Model 1; (b) Model 2 and (c) Model 3. The line is the exact result and the dots are the
simulational results using the procedure for Wang-Landau simulation \cite{caparica2}. The error bars are less than symbols.}
\label{lx}
\end{figure}

\begin{figure}[htbp]
\centering
\includegraphics[width=7.5cm,height=7.5cm]{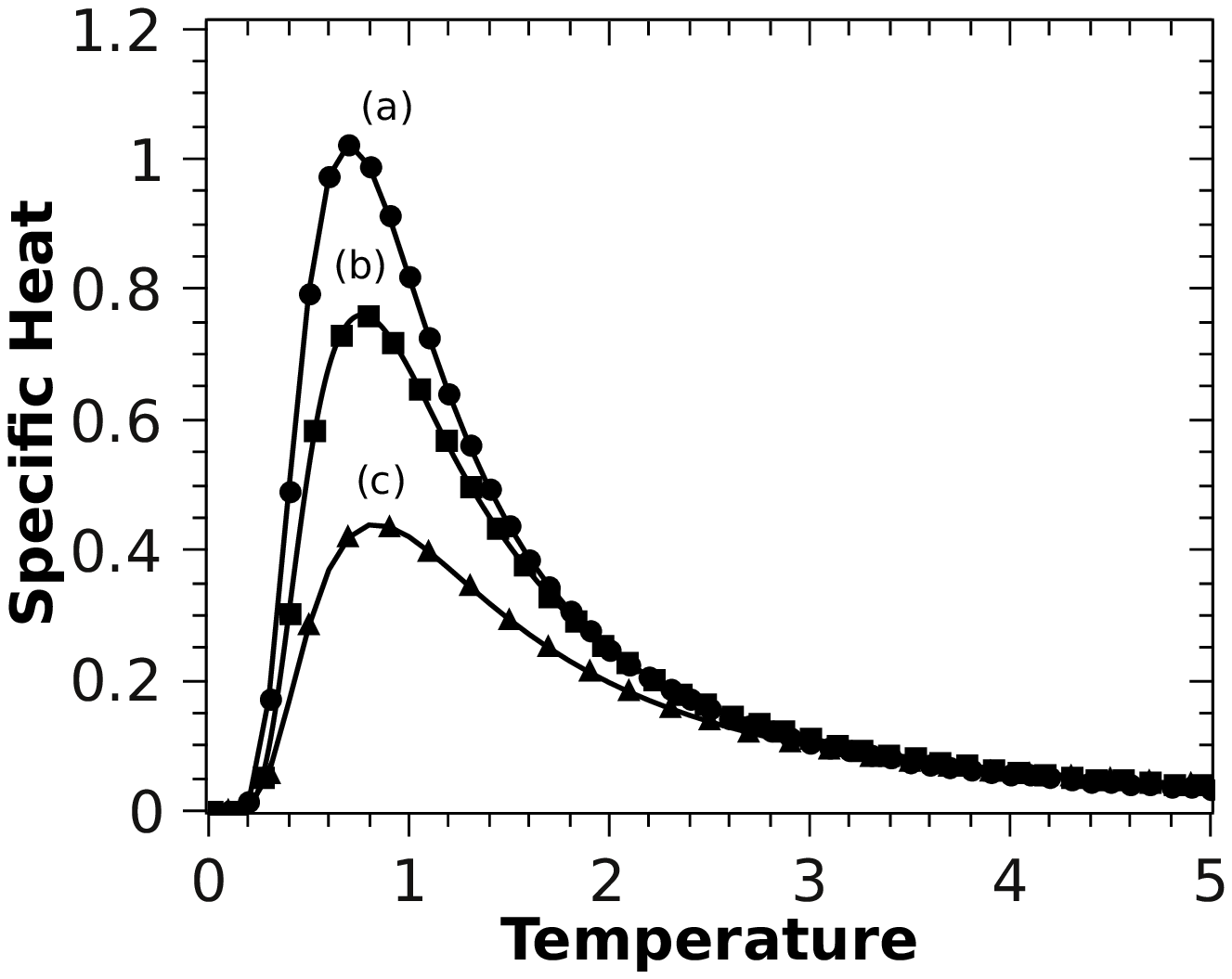}
\caption{Specific heat per monomer for polymer size $N=500$ for (a) Model 1; (b) Model 2 and (c) Model 3. The line is the exact result and the dots are the
simulational results using the procedure for Wang-Landau simulation \cite{caparica2}. The error bars are less than symbols.}
\label{cv}
 \end{figure}

In Figure \ref{ge} we plot in semi-logarithmical scale the density of states $g(E)$ for all the three models. Here we plot the results obtained from the simulations (symbols
in the figure) and also the exact theoretical results (continuous lines in the figure) for a polymer of $N=500$ monomers. In Figure \ref{lx} we plot the end-to-end distance
along the $x$-axis, $L_x$, as a function of the temperature. Here we rescale $L_x$ by the total number of monomers, $N=500$ in this case. We observed a very good agreement
between the simulation results (symbols in the figure) and the theoretical results (continuous lines in the figure), given by equations (\ref{LxN_1}), (\ref{LxN_2}), and
(\ref{LxN_3}), corresponding to model $1$, model $2$, and model $3$ respectively. In Figure \ref{cv} we plot the specific heat per monomers for polymers with $N=500$. The specific heat have a tail proportional to $1/T^{2}$ in the high temperature limit.

Finally, in Table I we present the location of peak of the specific heat for each model obtained by the Wang-Landau simulations
and compare with the results calculated with the exact density of states. The simulations were carried out for 100 independent
runs, adopting the $80\%$ flatness criterion. One can see that in the three cases the exact results fall into the error bars.
\vspace{.5cm}
\begin{center}
  \begin{tabular}{c c c}
\hline\hline
  Case    & \hspace{2cm}exact      &  \hspace{2cm}our results \\ [0.5ex]
\hline
  Model 1 & \hspace{2cm}0.70299027 &  \hspace{2cm}0.7043(23)   \\
\hline
  Model 2 & \hspace{2cm}0.75335362 &  \hspace{2cm}0.7513(23)  \\
  \hline
  Model 3 & \hspace{2cm}0.83180562 &  \hspace{2cm}0.8314(23) \\
\hline\hline
\end{tabular}
\vspace{.5cm}

\textbf{Table I:} \textsl{Temperatures of the peak of the specific heat from simulations, compared with the exact values. }
\end{center}

\section{Conclusions}
We have carried out Wang-Landau simulations of a simple polymer model which has exact solutions in both the microcanonical and the canonical ensembles. Here we considered three two-dimensional models: in the first model we allowed moves in all possible directions, in the second model the moves along the negative $x$-axis are forbidden, while in the last model we allowed only moves along the positive direction for both $x$ and $y$- axis. We have shown that updating the density of states only after each $N$ trial moves and halting the simulations when $f_{final}=f_{14}$ \cite{caparica2}, defined during the simulations, we obtain quite accurate results, compared with the available analytical exact results.  We have obtained a very good agreement between the simulations and the exact results also for the studied physical quantities: the end-to-end distance and the specific heat. As expected, due to the difference of the allowed directions of motion, in the limit of high temperatures the end-to-end distance has the highest value for the third model and the lowest for the first model.

\vspace{1.0cm}
\textbf{ACKNOWLEDGEMENT}

This work was partially supported by CNPq (Edital Universal) and FAPEAM (Programa Primeiros Projetos - PPP) (Brazilian Research Agency).

\vspace{10.0cm}

\vspace{0.1cm}

\vspace{0.1cm}

\vspace{0.1cm}

\end{document}